\documentclass[epj,referee]{svjour}
\usepackage{graphics}
\begin{document}
\title{The market efficiency in the stock markets}
\author{Jae-Suk Yang\inst{1}\and Wooseop Kwak\inst{1}\and Taisei Kaizoji\inst{2}\and  In-mook Kim\inst{1}
\thanks{\emph{e-mail:} imkim@korea.ac.kr}%
}                     
\offprints{In-mook Kim}          
\institute{Department of Physics, Korea University, Seoul 131-701,
Republic of Korea \and Division of Social Sciences, International
Christian University, Osawa, Mitaka, Tokyo 181-8585, Japan}
\date{Received: \today / Revised version: date}
%
\abstract{ We study the temporal evolution of the market
efficiency in the stock markets using the complexity, entropy
density, standard deviation, autocorrelation function, and
probability distribution of the log return for Standard and Poor's
500 (S\&P 500), Nikkei stock average index, and Korean composition
stock price index (KOSPI). Based on a microscopic spin model, we
also find that these statistical quantities in stock markets
depend on the market efficiency.
} 

\maketitle
\section{Introduction}
\label{intro} Econophysics is one of the most active fields in
interdisciplinary research
\cite%
{eguiluz,krawiecki,chowdhury,takaishi,kaizoji02,kaizoji04,jbpark,palagyi,kaizoji01,kaizoji03,matal,yang,silva}%
. Time series analysis and agent based modelling have been studied
by many researchers. There are many methodologies to analyze the
financial time series. Observing probability distribution
functions (FDFs) of log return is one of the simplest and the most
popular methods. Many research papers about
PDFs of log return for stock markets have already been published \cite%
{kaizoji03,matal,yang,silva,stanley1,mccauley}. The different
characteristics between mature markets and emerging markets
\cite{matal}, market efficiency \cite{yang}, and the relation
between shape of PDFs and time lags \cite{silva} are studied using
PDFs. Also it is used to distinguish between bubble and
anti-bubble \cite{kaizoji01,kaizoji03}.

Another method is computational mechanics \cite{shalizi1}.
Computational
mechanics has been studied various fields of science \cite%
{hanson,shalizi,crutchfield}, and it is applied to analyze the
stock market \cite{jbpark}. Computational mechanics is available
to analyze complexity and structure quantitatively by finding
intrinsic causal structures of time series \cite{clarke}.

Agent based modelling has been widely used in social science and
econophysics to construct artificial social and economic systems.
Agent
based models in econophysics are constructed using agents clustering \cite%
{eguiluz}, Ising-like spin model \cite{krawiecki,yang}, and
Potts-like model
\cite{takaishi}. Variation of PDFs shapes by traders' characteristics \cite%
{kaizoji03} and information flow \cite{yang}, and speculative
activity explaining bubbles and crashes in stock market
\cite{kaizoji02} have been simulated by agent based model.

In this paper, we analyze the time series of Standard and Poor's
(S\&P 500), Nikkei stock average index, and Korean composition
stock price index (KOSPI) by time evolution of statistical
measures such as PDFs of log return, autocorrelation function,
complexity, entropy density, and scaling properties of the
standard deviation of log return. Moreover, we construct the stock
market using microscopic spin model to simulate above time series
results.

\section{Empirical data and analysis}

We use the S\&P 500 data mainly for the period from 1983 to 2006.
Japanese data for the period from 1997 to 2005 and Korean data for
the period from 1992 to 2003 are also used to support and confirm
the results from S\&P 500. The data resolution is high frequency
(1 minute) data, and we use only intra-day returns to exclude
discontinuity jumps between the previous day's close and the next
day's open price due to the overnight effects. The price return is
defined as
\begin{equation}
S(t)\equiv \log Y(t+\Delta t)-\log Y (t),  \label{eq:return}
\end{equation}
where $Y(t)$ is the price at time $t$ and $\Delta t$ is the time
lag.

\subsection{Probability distribution and autocorrelation}

The distribution of price changes are identified as non-Gaussian
\cite{yang,silva,stanley1,mccauley,vicente}. Especially, when the
PDF has the power law tail, the exponent of power at tail part can
be gotten from the PDF. That exponent is called as tail index.

\begin{figure}[tbph]
\begin{center}
\mbox{
\resizebox{0.75\columnwidth}{!}{%
  \includegraphics{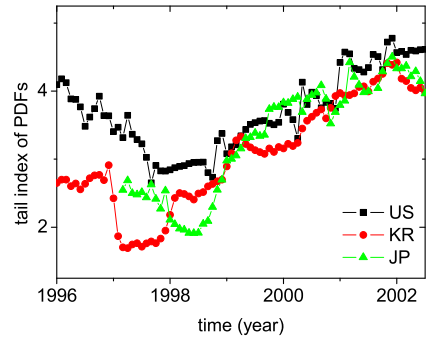}}
}
\mbox{
\resizebox{0.75\columnwidth}{!}{%
  \includegraphics{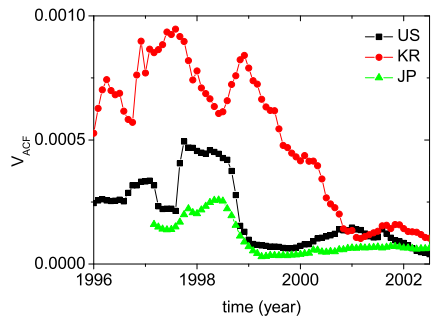}}
}
\end{center}
\caption{(a) Temporal evolution of tail index and (b) variance of
autocorrelation function for the S\&P 500, the Nikkei stock index,
and the KOSPI.} \label{fig1}
\end{figure}

Fig. \ref{fig1}a shows temporal evolution of tail index in PDFs
for S\&P 500, Nikkei stock index, and KOSPI. Tail index of PDFs
increases from around 2 to above 4 as time passes. In 2000s, the
shape of PDF becomes narrower and the tail part becomes thinner,
while PDF has fatter tail and the slope of tail part is more steep
in 1990s. Autocorrelation function is defined as follows:
\begin{equation}
R(\tau) = \frac{<S(t)S(t+\tau)>}{\sigma^{2}},
\end{equation}
where $\sigma$ is a standard deviation of $S(t)$. Moreover, the
variance of autocorrelation function is defined as follows:
\begin{equation}
V_{ACF} = <R(\tau)^{2}>.
\end{equation}
Fig. \ref{fig1}b shows the temporal evolution of variance of
autocorrelation function. The increasing tendency for tail index
is reverse to it for variance of autocorrelation function. We can
guess that the reason why probability distributions of log return
are changed is related to autocorrelation of log return time
series.

Though the tendency is same for three stock markets, the value of
tail index for the S\&P 500 is larger than it for the KOSPI and
$V_{ACF}$ for S\&P 500 is smaller than it for the KOSPI in the
1990s.

\subsection{Scaling property of standard deviation}
\begin{figure}[tbph]
\begin{center}
\resizebox{0.75\columnwidth}{!}{%
  \includegraphics{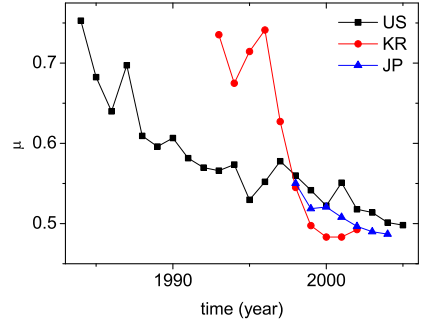} }

\end{center}
\caption{Temporal evolution of scaling properties of the standard
deviation of log return.} \label{fig2}
\end{figure}

We investigate the long range memory of log return by observing
the time evolution of scaling properties in the standard deviation
of log return \cite{palagyi}. The standard deviation of log return
is defined as
\begin{equation}
\sigma (\Delta t)=\frac{\sqrt{\sum_{i=1}^{n}\left( \log
{Y(t_{i}+\Delta t)-\log {Y(t_{i})}}\right) ^{2}}}{\sqrt{n-1}},
\label{eq:stdDev}
\end{equation}%
as a function of the time lag $\Delta t$. The relation between
standard deviation and time lag is as follows:
\begin{equation}
\sigma (\Delta t)\sim {\Delta t}^{\mu }.  \label{eq:rel}
\end{equation}%
When $\mu $ is larger than 0.5, the time series has long range
correlation, while long range anticorrelation when $\mu <0.5$.
There is no correlation at $\mu =0.5$ and strength of correlation
(or anticorrelation) is proportional to the difference between
$\mu $ and 0.5. Fig. \ref{fig2} shows the temporal evolution of
scaling properties of the standard deviation of log return. The
value of $\mu $ decreases to around 0.5. Until the mid 1990s, time
series of stock market index has strong long range correlation.
However, long range correlation practically disappears in 2000s.

In spite of the same tendency for temporal evolution of $\mu$, the
S\&P 500 is more close to 0.5 than the KOSPI in the 1990s.

\subsection{Entropy density and statistical complexity}

We also analyze financial time series using computational
mechanics to find the statistical complexity and the entropy
density. In order to calculate statistical complexity, we used
causal-state splitting reconstruction (CSSR) algorithm \cite{shalizi1} to model $\epsilon$%
-machine of the stock markets.

To calculate the entropy density and the statistical complexity,
we should symbolize the time series as follows:
\begin{equation}
F(t) \equiv \theta(Y(t+\Delta t)-Y(t)),
\end{equation}
where $\theta(x)$ is a Heaviside step function. Then the original
data $Y(t)$ are changed into the binary time series $F(t)$ with a
countable set $A=\{0, 1\}$. $F(t)$ is 0 (or 1) when the next index
has decreased (or increased).

Claude Shannon suggested the entropy of a discrete random variable
$X$ with a probability function $P(x)$ \cite{sloane} as follows:
\begin{equation}
H[X] = - \sum_{x} P(x)\log_{2} P(x).
\end{equation}
Let $A$ be a countable set of symbols of time series and let $S$
be a random variable for $A$, and $s$ is its realization. If a
block of string with $L$ consecutive variable is denoted as $S^L =
S_{1} , ..., S_{L}$, then Shannon entropy of length $L$ is defined
as
\begin{equation}
H[X] = - \sum_{s_{1} \in A}\cdots \sum_{s_{L}\in A}
P(s_{1},...,s_{L})\log_{2} P(s_{1},...,s_{L}).
\end{equation}
Also entropy density for the finite length $L$ is define as
\begin{equation}
h_{\mu}(L) \equiv H(L)-H(L-1),
\end{equation}
as a function of block length $L$ where $L=1, 2, 3, \cdots$.
Entropy density is more useful because it is normalized quantity
while $H(L)$ also increases as $L$ increases.

In next, to calculate statistical complexity $\epsilon $-machine
has to be
defined. An infinity string $S^{\hspace{-2.5mm}\raisebox{1mm}{$%
\leftrightarrow$}}$ can be divided into two semi-infinite parts
such as a
future $S^{\hspace{-2.5mm}\raisebox{1mm}{$\rightarrow$}}$ and a history $S^{%
\hspace{-2.5mm}\raisebox{1mm}{$\leftarrow$}}$. A causal state is
defined as a set of histories that have the same conditional
distribution for all the futures. $\epsilon $ is a function that
maps each history to the sets of histories, each of which
corresponds to a causal state:
\begin{eqnarray}
\epsilon (s^{\hspace{-2.5mm}\raisebox{1mm}{$\leftarrow$}})
&=&\{s^{\prime
\hspace{-2.5mm}\raisebox{1mm}{$\leftarrow$}}\mid P(\vec{S}^{L}=\vec{s}%
^{L}\mid S^{\hspace{-2.5mm}\raisebox{1mm}{$\leftarrow$}}=s^{\hspace{-2.5mm}%
\raisebox{1mm}{$\leftarrow$}})=P(\vec{S}^{L}=\vec{s}^{L}\mid S^{\hspace{%
-2.5mm}\raisebox{1mm}{$\leftarrow$}}=s^{\prime \hspace{-2.5mm}%
\raisebox{1mm}{$\leftarrow$}}),  \nonumber \\
\vec{s}^{L} &\in &\vec{S}^{L},s^{\prime \hspace{-2.5mm}\raisebox{1mm}{$%
\leftarrow$}}\in
S^{\hspace{-2.5mm}\raisebox{1mm}{$\leftarrow$}},L\in
\mathbf{Z^{+}}\}.
\end{eqnarray}

The transition probability $T_{ij}^{(a)}$ denotes the probability
of generating a symbol $a$ when making the transition from state
$S_i$ to state $S_j$ \cite{shalizi02,feldman}.

The combination of the function $\epsilon$ from histories to
causal states
with the labelled transition probabilities $T_{ij}^{(a)}$ is called the $%
\epsilon$-machine \cite{shalizi02}, which represents a
computational model underlying the given time series.

Given the $\epsilon$-machine, statistical complexity is defined as
\begin{equation}
C_{\mu} \equiv - \sum_{\{S_i \}} P (S_i ) \log _{2} P (S_i ).
\end{equation}

\begin{figure}[tbph]
\begin{center}
\mbox{
\resizebox{0.75\columnwidth}{!}{%
  \includegraphics{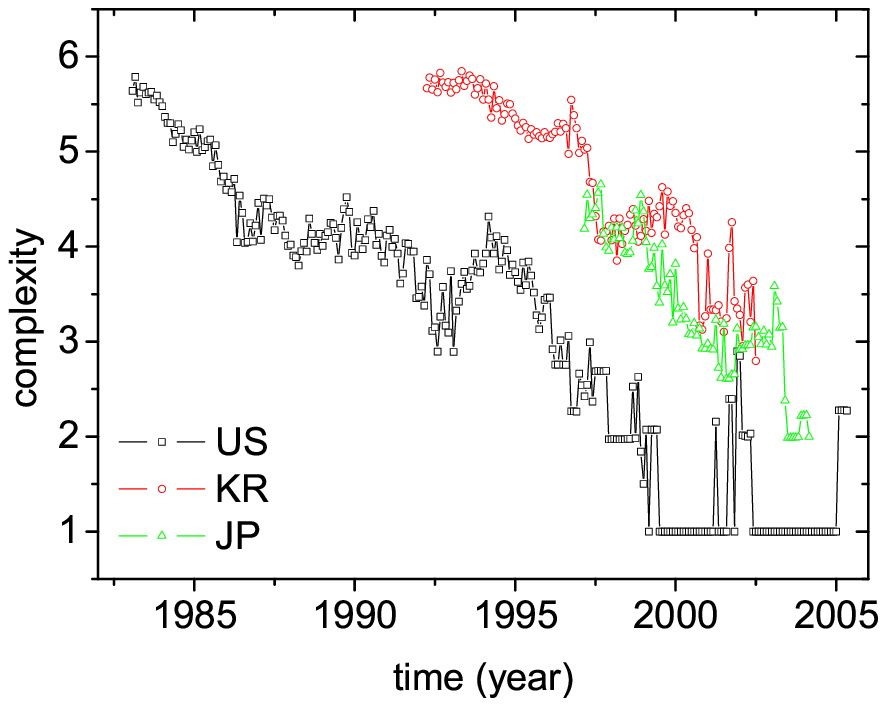} }
} \mbox{
\resizebox{0.75\columnwidth}{!}{%
  \includegraphics{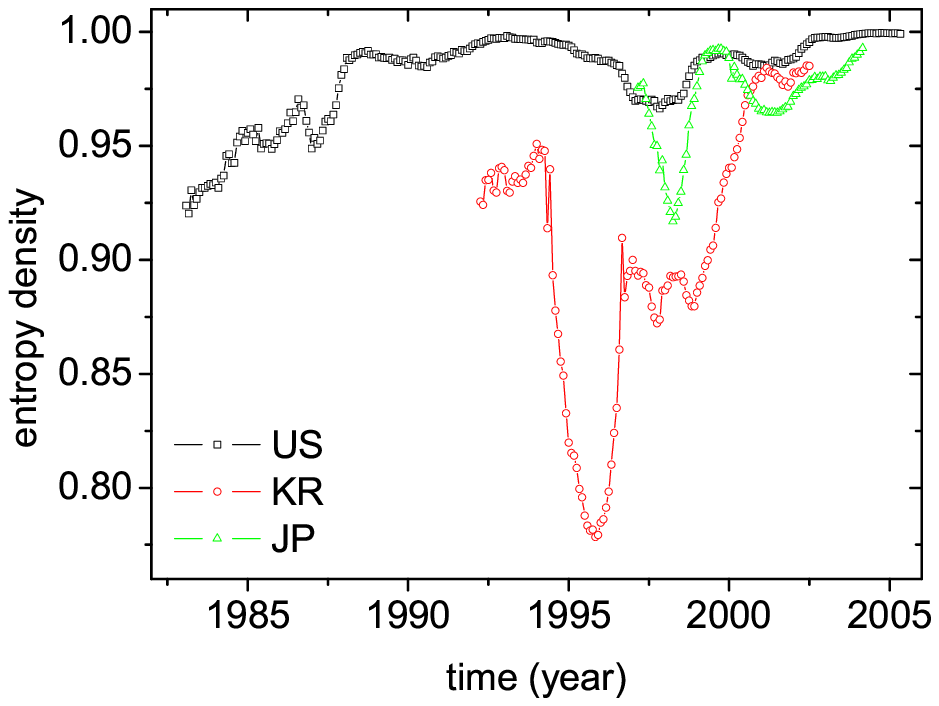} }
}
\end{center}
\caption{Temporal evolution of (a) statistical complexity and (b)
entropy density.} \label{fig3}
\end{figure}

Fig. \ref{fig3} shows temporal evolution of statistical complexity
and entropy density. Statistical complexity decreases and entropy
density increases in all three markets as time passes.

Statistical complexity is around 0 when time series has regular
pattern or it is totally random. To clarify whether the time
series is random or regular, the entropy density is needed. Time
series is totally random when entropy density is around 1 and
entropy density is 0 when time series has periodic pattern because
it is a measure of disorder. So we can find out that the time
series of stock markets is getting more randomly and the patterns
in the time series almost disappear in 2000s. Also the values of
complexity and entropy density for the S\&P 500 are different from
them for the KOSPI, though the inclination is same.

\section{Model and results}

We constructed the microscopic model of many interacting agents to
simulate the variation of some statistical characteristics for the
stock price time series by modifying microscopic spin model
\cite{krawiecki}. The number of
agents is $N$, and we consider $i=1,2,\dots ,N$ agents with orientations $%
\sigma_{i} (t) = \pm 1$, corresponding to the decision to buy
($+1$) and sell ($-1$) stock at discrete time-steps $t$. The
orientation of agent $i$ at the next step, $\sigma_{i}(t+1)$,
depends on the local field:
\begin{equation}
I_{i}^{pri} (t) = \frac{1}{N} \sum _{j} A_{ij} (t) \sigma_{j} (t)
+ h_{i} (t),
\end{equation}
where $A_{ij} (t)$ represent the time-dependent interaction
strength among agents, and $h_{i} (t)$ is an external field
reflecting the effect of the environment. The time-dependent
interaction strength among agents is $A_{ij} (t) = A\xi (t) + a
\eta_{ij}(t)$ with $\xi(t)$ and $\eta_{ij} (t)$ determined
randomly in every step. $A$ is an average interaction strength and
$a$ is a deviation of the individual interaction strength. The
external field reflecting the effect of the environment is
$h_{i}=h \zeta_{i} (t)$, where $h$ is an information diffusion
factor, and $\zeta_{i}(t)$ is an event happening at time $t$ and
influencing the $i$-th agent.

From the local field determined as above, agent anticipates log
return of stock index as follows:
\begin{equation}
x_{i}^{exp}(t)=\frac{2}{1+e^{-2I_{i}^{pri}(t)}}-1.
\end{equation}
So, the local field on agent can be represented as follows:
\begin{equation}
I_{i} (t)= I_{i}^{pri} (t) +\alpha \left( x(t-1)-x_{i}^{exp}(t-1)
\right),
\end{equation}
where $\alpha$ is degree of adjustment. When $\alpha=0$, agents
determine their opinion from $I_{i}^{pri}$, while agents determine
their opinion from the price or log return of previous step as
well as information flowed into the market when $\alpha$ is
non-zero. For instance, in case positive $\alpha$ and $x(t-1)
> x_{i}^{exp}(t-1)$, agents determine their opinion by adding
the difference between the market price changes and the
anticipated price changes at the previous step. On the contrary,
in case $x(t-1) < x_{i}^{exp}(t-1)$, agents subtract the
difference from $I_{i}^{pri}$ to adjust their inexact information.
By this way agents refer to past performance, while agents act by
fundamental expressed by $I_{i}^{pri}$ in case of $\alpha = 0$.

From the local field determined as above, agent opinions in the
next step are determined by:
\begin{equation}
\sigma_{i} (t+1) =\left\{
\begin{array}{ll}
+1 & \mbox{with
probability $p$} \\
-1 & \mbox{with probability $1-p$}%
\end{array}
\right.,
\end{equation}
where $p=1/(1+exp\{-2I_{i}(t)\})$. In this model, price changes
are:
\begin{equation}
x(t) = \frac{1}{N} \sum \sigma_{i}(t).
\end{equation}

\begin{figure}[tbph]
\begin{center}
\mbox{
\resizebox{0.75\columnwidth}{!}{%
  \includegraphics{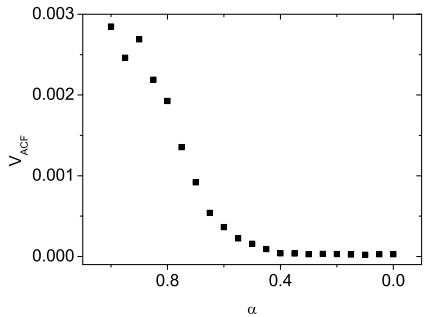} }
} \mbox{
\resizebox{0.75\columnwidth}{!}{%
  \includegraphics{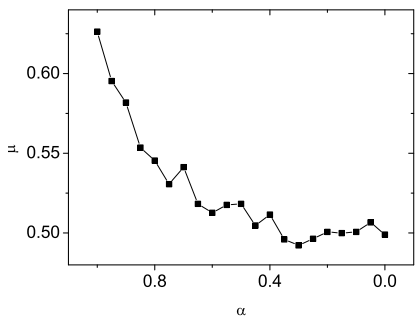} }
}
\end{center}
\caption{(a) Variance of autocorrelation function for various and
(b) Scaling exponents of standard deviation for various $\alpha$.}
\label{fig4}
\end{figure}

Fig. \ref{fig4}a shows variance of autocorrelation function for
various $\alpha$. As $\alpha $ decreases, the tail is getting
thinner and thinner, and the strength of autocorrelation is
reduced. Moreover, scaling exponents of standard deviation go to
0.5 as $\alpha$ decreases [see Fig. \ref{fig4}b]. The generated
time series has long range correlation for larger $\alpha$, and
Correlation is almost disappeared for small value of $\alpha$.

\begin{figure}[tbph]
\begin{center}
\mbox{
\resizebox{0.75\columnwidth}{!}{%
  \includegraphics{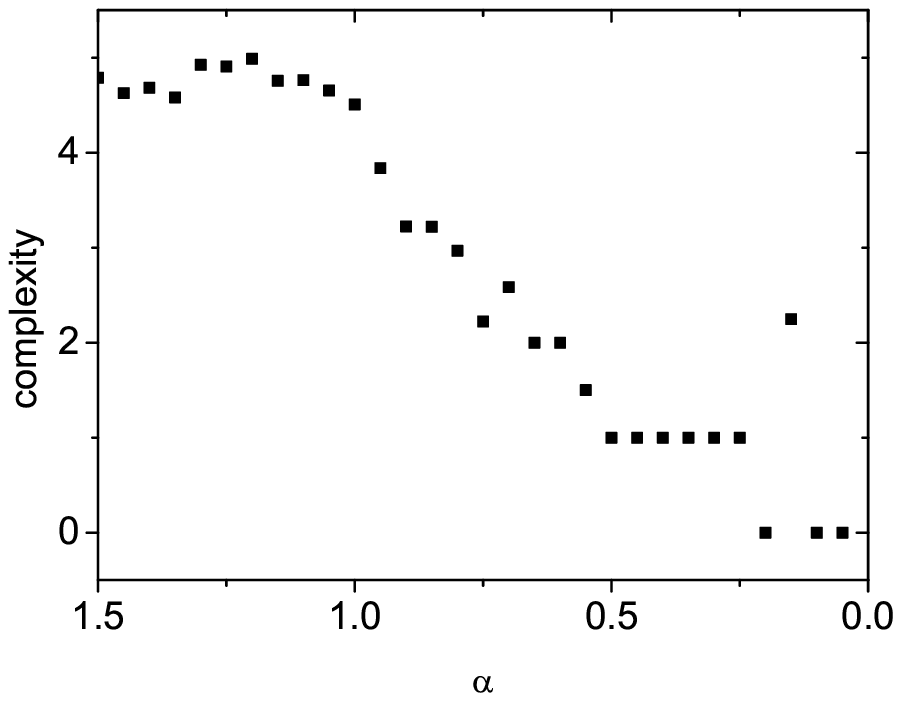}} }
\mbox{
\resizebox{0.75\columnwidth}{!}{%
  \includegraphics{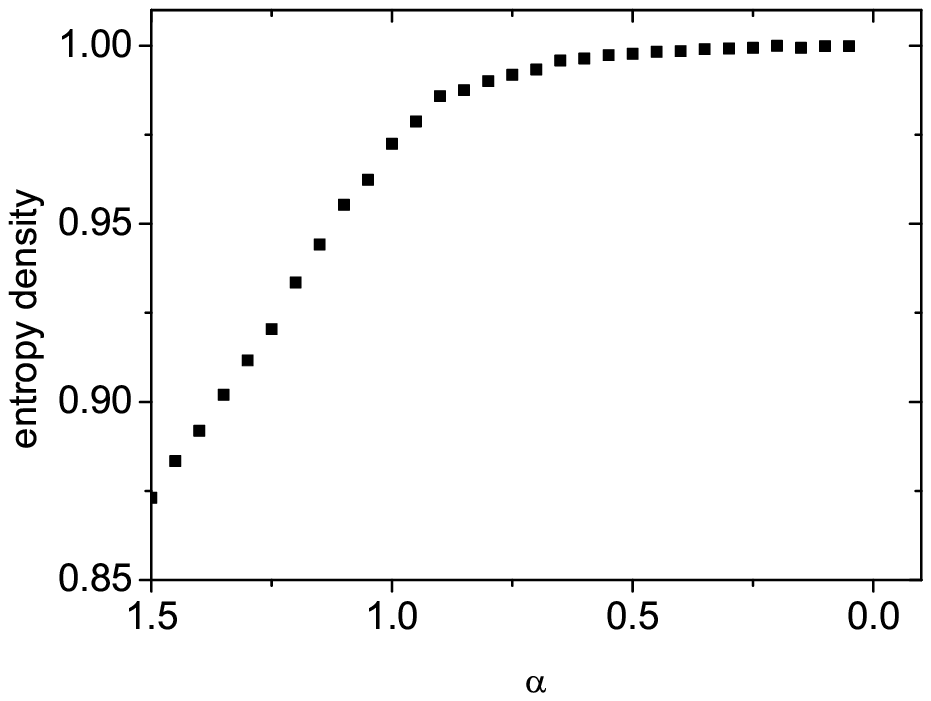}} }
  \mbox{
\resizebox{0.75\columnwidth}{!}{%
  \includegraphics{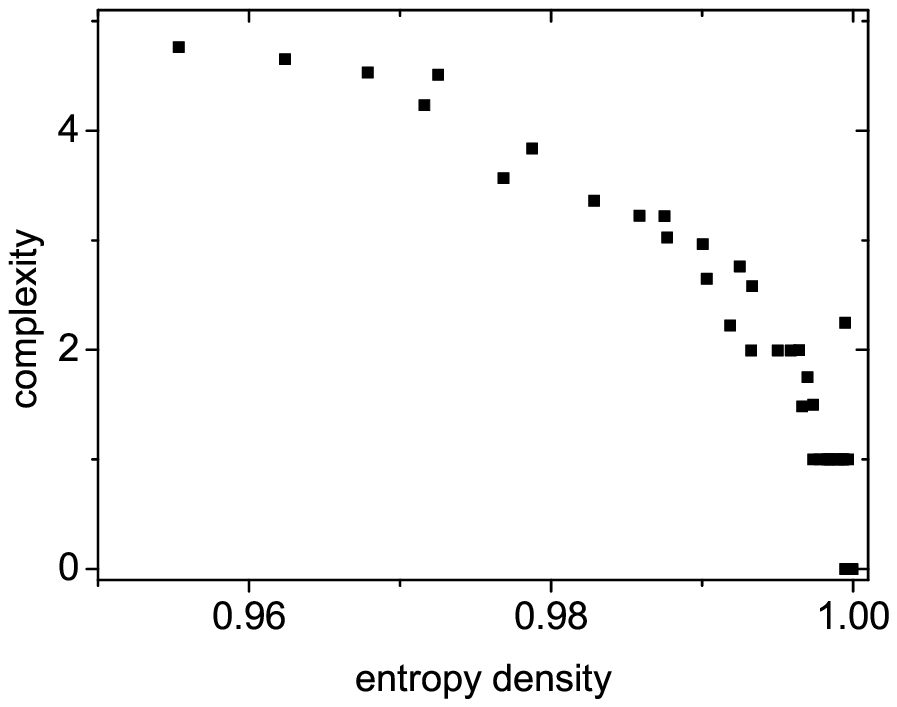}} }
\end{center}
\caption{(a) Statistical complexity, (b) entropy density for
various $\alpha$, and (c) the relation between entropy density and
statistical complexity.} \label{fig6}
\end{figure}

In Fig. \ref{fig6}, we can confirm the tendency of statistical
complexity and entropy density for various $\alpha$. As $\alpha$
decreases, statistical complexity decreases and entropy density
increases. From the statistical complexity, the pattern in time
series is getting simpler or the degree of randomness of times
series is larger for smaller $\alpha$. What entropy density is 1
means the time series is practically random. Fig. \ref{fig6}c is
the relation between entropy density and statistical complexity.
From this relation we can distinguish if the time series is random
or regular.

\section{Conclusions}

We analyze the time series of stock index of U. S., Japan, and
Korea using some statistical measures and simulate them by
microscopic agent based spin model.

Time series has a fat tail in log return distribution and a tail
index is increased as time passes to present. Existence of pattern
in the financial time series can be confirmed by autocorrelation
function, entropy density and complexity. As time goes from past
to present, entropy density is increased and complexity is
decreased. Also autocorrelation is decreased. From these results,
the relation between present data and past data is decreasing and
the pattern in stock log return data disappears. 

In the spin model, when $\alpha$ is non-zero, traders adjust their
opinion using the difference between their anticipated prices and
real market prices, and they anticipate price changes of next step
with adjusted information. In the past, the speed of information
is slower and market is less efficient, so adjusting behavior is
more effective and active in the same time interval compare to
present. Therefore, the past market corresponding to higher
$\alpha$ has long range correlation and vice versa.

$I_{i}^{pri}(t)$ is generated randomly because its elements are
random variables while $\alpha \left(
x(t-1)-x_{i}^{exp}(t-1)\right) $ provides regularity to $I_{i}(t)$
because effect of this term remains for a while like Markov chain.
When $\alpha $ is
0, entropy density is almost 1 and complexity is 0 because time series for $%
\alpha =0$ are almost random. As $\alpha $ is increased, entropy
density is decreased and complexity is increased because the
pattern is generated in the time series.

The reason why these changes occur is that speed of information
flow is becoming fast by the development of infra for
communication such as high speed internet, mobile communication
and broadcasting systems. So market has become more efficient. By
the efficient market hypothesis (EMH), the speed of information is
so fast that agents can not gain profit by superiority of
information.

We would like to thank Hang-Hyun Jo for helpful discussions. This
work is supported by the Second Brain Korea 21 project and also by
the Grant No. R01-2004-000-10148-1 from the Basic Research Program
of KOSEF.

\end{document}